\documentclass[aps,prl,superscriptaddress,showpacs,endfloats,preprint]{revtex4}

\usepackage{graphicx}	

\begin{document}

\title{Prior to ``Quark Matter 2006'' predictions within retarded jet absorption scenario at RHIC}

\author{V.S. Pantuev\\
University at Stony Brook, Stony Brook, NY 11794-3800\\
On leave from Institute for Nuclear Research Russian Academy of Sciences, Moscow}   \noaffiliation


\begin{abstract}

Predictions for some experimental physical observables in nucleus-nucleus collisions at 
RHIC energies are presented. I  utilize the previous suggestion that 
the retarded, by time about 2-3 fm/c, jet absorption in opaque core is a natural explanation of many 
experimental data. This assumption is applied only to the particles with high 
transverse momentum above 4 GeV/c resulting from parton fragmentation. 
I calculate nuclear modification factor $R_{AA}$, $R_{AA}$ in- and out of 
the reaction plane, 
azimuthal anisotropy parameter $v_2$, jet suppression $I_{AA}$ for the away side jet and its
dependence versus the reaction plane orientation. The systems 
under consideration are Au+Au, Cu+Cu, at 200 GeV and 62 GeV.
Most of numbers are  predictions prior to the QM2006.

\end{abstract}



\maketitle

At QM2006 conference we expect that many new experimental data will be shown. I already get few requests 
from different groups to calculate some experimentally observed parameters. To be fair, I 
present all these numbers prior to the conference.

In previous papers~\cite{mypaper, paper2} I propose a simple model to explain 
experimental data on the angular dependence of the nuclear modification factor $R_{AA}$ at high 
transverse momentum  
in the reaction plane. The model has one free parameter $L$$\simeq$2 fm to describe the the thickness of 
the corona area with no absorption and was adjusted to fit the experimental data on  AuAu collisions at 
centrality 50-60$\%$. The model nicely 
describes the $R_{AA}$ dependence for all centrality classes. I extract the second Fourier 
component amplitude, $v_2$, for high pt particle azimuthal distribution and found $v_2$ should 
be at the level of 11-12$\%$ purely from the geometry of the collision with particle absorption 
in the core. At that time I made a prediction for $R_{AA}$ in Cu+Cu collisions at 200 GeV which,
as later was found, is in very good agreement with experimental data. Physical 
interpretation of the parameter $L$ could be that it is actually retarded jet absorption caused 
by the plasma formation time $T=L/c$ $\simeq$ 2 fm/c, or at least non-trivial response of strongly 
interacting plasma  to fast moving color charge.

This value of formation time is true for the dilute corona 
region where density is small. For more central interaction region it should be smaller, probably 
changing like mean distance between interaction points or $1/\sqrt(\rho(x,y))$, where $\rho(x,y)$ is 
two-dimentional (projected into $x-y$ plane) density of $N_{binary}$ or $N_{part}$. 
Extrapolation of  $T$ in a such way gives values of the order of 
0.7-1 fm/c in the center of collision zone. 

As usual, nuclear modification factor $R_{AA}$ is defined as:
\begin{equation}
R_{AA}(p_T) 
= \frac{(1/N_{evt}) \; d^{2}N^{A+A}/dp_T d\eta }
{(\langle N_{binary} \rangle/\sigma^{N+N}_{inel}) \; d^{2}\sigma^{N+N}/dp_T d\eta},
\label{eq:RAA_defined}
\end{equation}
where $\langle N_{binary} \rangle$ is a number of binary nucleon-nucleon collisions at particular 
centrality class.

\begin{figure}[thb]
\includegraphics[width=1.0\linewidth]{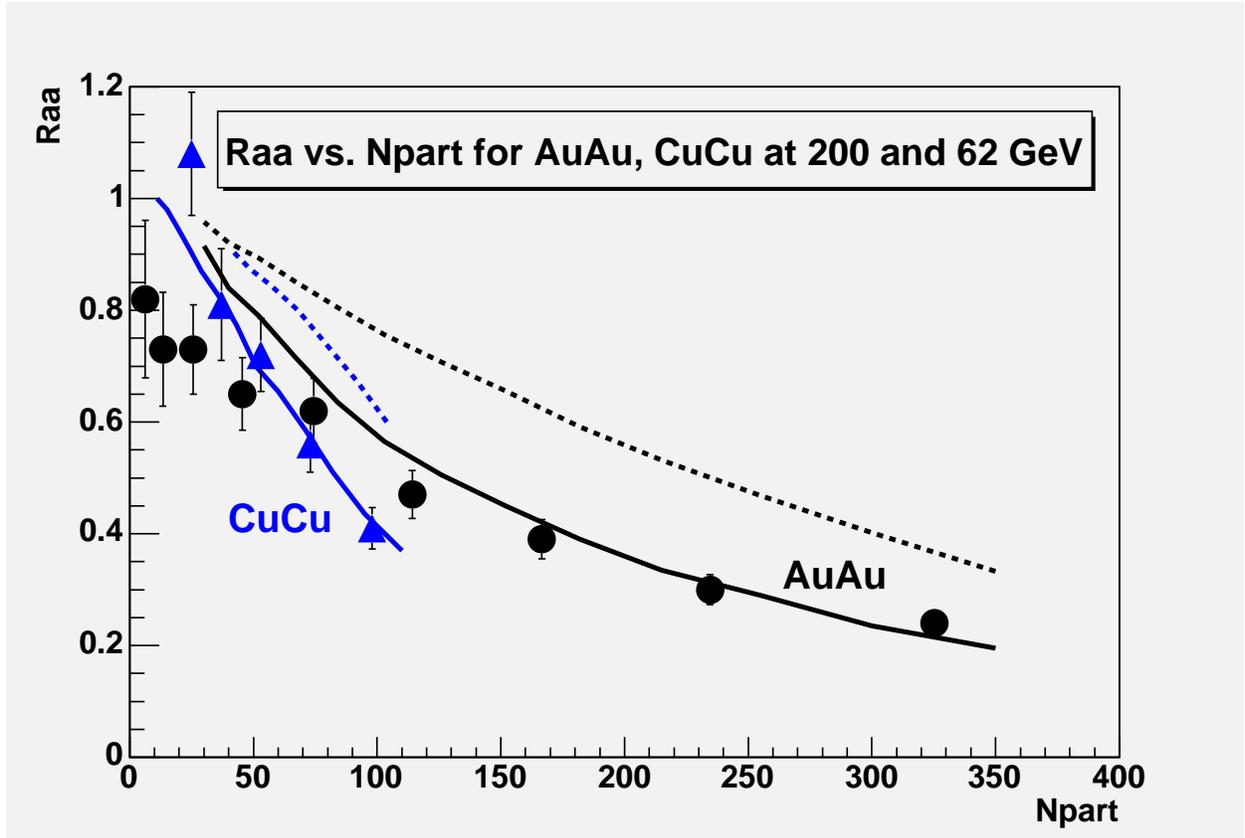}
\caption{\label{fig:Raa}  Calculated $R_{AA}$  for Au+Au and Cu+Cu collisions 
at 200 GeV and 62 GeV versus the number of
participant nucleons, $N_{part}$. Solid curves are for 200 GeV, dashed are for 62 GeV. 
The circles are experimental $\pi^0$ data for 200 GeV Au+Au collisions integrated for 
$p_T$$\ge$4 GeV/c~\cite{pi0}. The triangles are data for 200 GeV Cu+Cu collisions of  
$p_T$$\ge$7 GeV/c~\cite{maya}. Only statistical errors are shown.}
\end{figure}

I use a  Monte Carlo simulation of nucleus-nucleus collisions based on 
the Glauber approach with Woods-Saxon density distribution. Assume that jets are produced in accordance with 
the $N_{coll}$ distribution in the $x-y$ transverse plane. I neglect longitudinal expansion for the first 
2-3 fm/c. Let all jets fly in all possible directions (some of them will move from the interaction 
zone determined by 
the envelop of Woods-Saxon  radii, some will go into this zone) for some time $T$. If after that time jet will 
be still inside the interaction zone, it will be completely absorbed. If not, jet leaves the zone without 
interaction. We found from experimental data that this time should be about 2.3 fm/c at 200 GeV and 3.5 fm/c 
at 62 GeV center of mass beam energy. Absorption can not happen instantly (or in the infinitely thin layer), 
plus there should be statistical fluctuations. To take these into account I smooth 
the cut edge: an arbitrary weight function in the form of a Fermi distribution  with diffuseness 
parameter $a$ was 
applied:

\begin{equation}
weight(l) 
= \frac{1}{1+exp((T-l)/a)}.
\label{eq:weight}
\end{equation}
The value of this function changes from 0.1 to 0.9 at $\Delta$l=4.4*$a$ around $T$.  
Parameter $a$ varied from 0.01 to 0.5 fm/c. It was found that the results for $R_{AA}$ and $v_2$ do not change much 
for this parameter range. By default it was chosen to be $a$=0.2 fm/c. The results of calculation 
for $R_{AA}$ are 
presented in Fig.~\ref{fig:Raa}. All values for $R_{AA}$ and $v_2$ are presented in tables at 
the bottom of this preprint. It is worth to mention, values of $R_{AA}$ and $v_2$ 
for mesons and baryons at high $p_T$ should be the same. Obviously, 
$R_{AA}$ is  flat versus $p_T$ 
if no other physics process is ``switched on'' at high $p_T$, except retarded absorption 
in opaque core considered here.

For di-jets, another parameter, $I_{AA}$, was introduced in Ref.~\cite{starIaa}. For away side jet $I_{AA}$ is the ratio of 
particle yield per trigger in the particular momentum range of trigger and associate particles, to the similar 
yield in p+p collisions. To calculate this value I perform the same procedure like for $R_{AA}$, except that in addition 
the location of away side jet after time $T$ should be taken into account. Away side jet may not go exactly 
in opposite  relative to the trigger jet direction, but is distributed around 180 degree angle. 
It introduces another critical parameter for the calculation, which should be taken from the experimentally 
observed width  of the away side jets in azimuthal angle. I use simple Gausse distribution around 180 degrees for 
the away side jet.  
Depending on the momentum range of the  associate particle, width of the away side jet varies from 0.5 radians 
for the low momentum range to 0.2 radians at momentum above 5-6 GeV/c.

\begin{figure}[thb]
\includegraphics[width=1.0\linewidth]{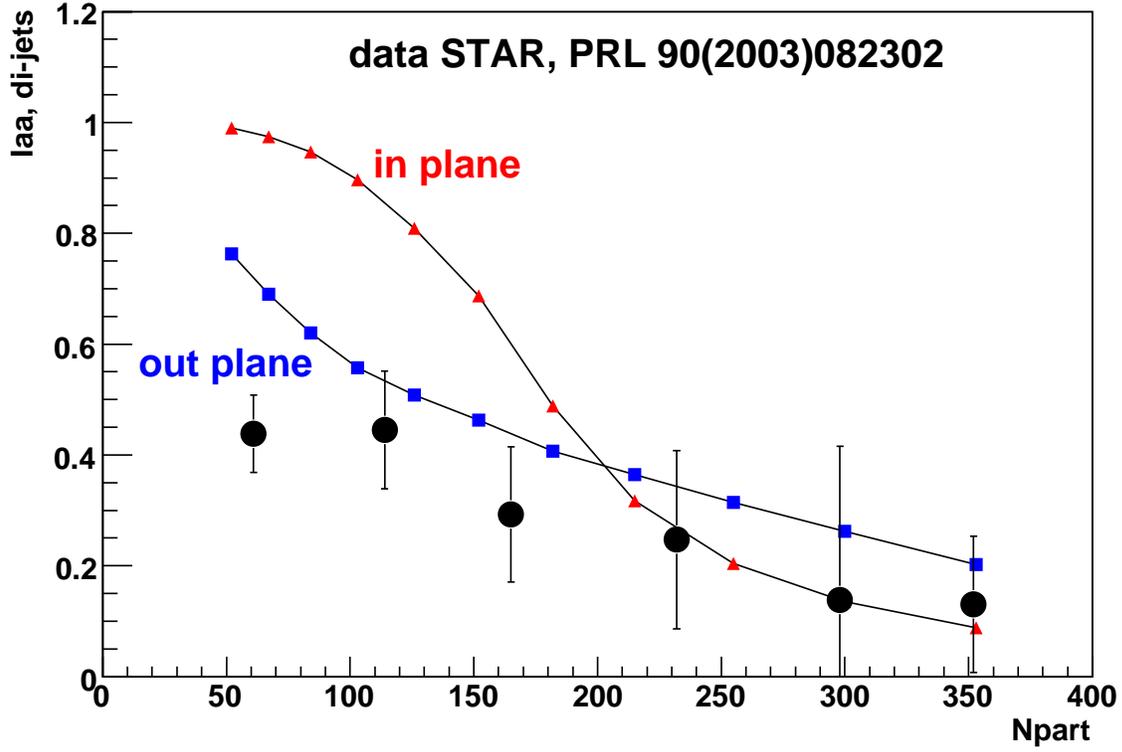}
\caption{\label{fig:Iaa}  Calculated $I_{AA}$ in- and out of plane for Au+Au collisions 
at 200 GeV versus the number of
participant nucleons, $N_{part}$. Width of the away side jet is $\sigma$=0.22 radians. 
Small triangles are estimation for in plane,  squeres- 
for out of plane. 
The circles with error bars are data from~\cite{starIaa}.}
\end{figure}

\begin{figure}[thb]
\includegraphics[width=1.0\linewidth]{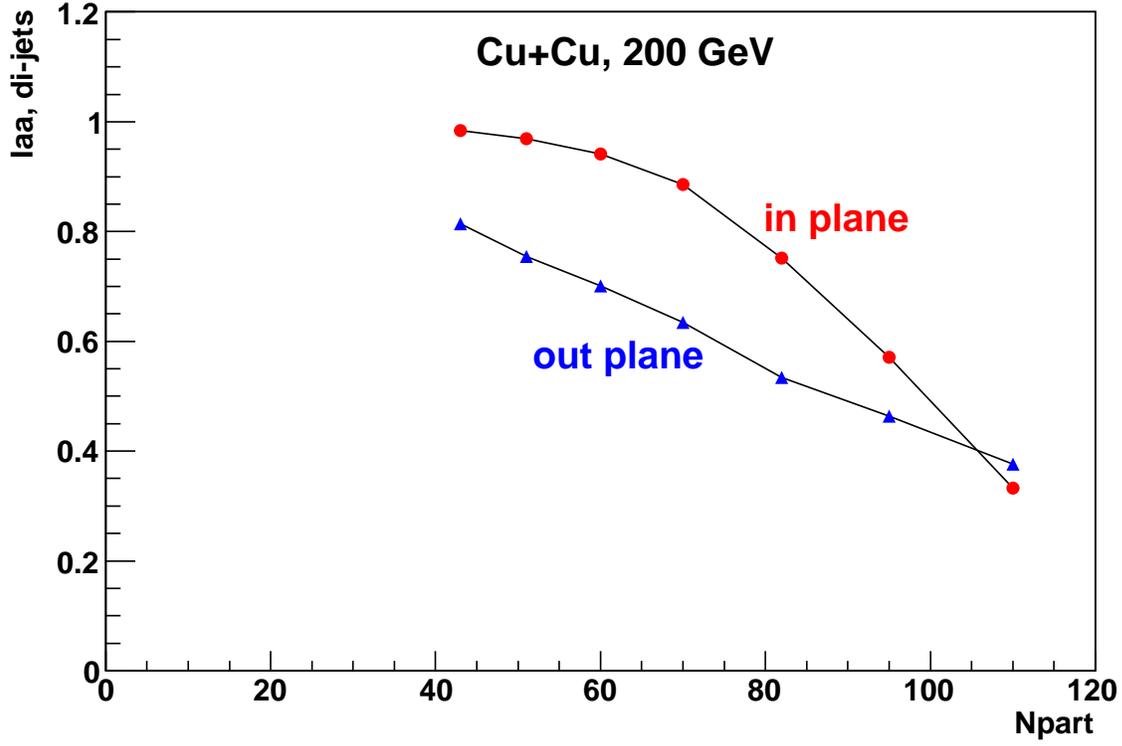}
\caption{\label{fig:IaaCu}  Calculated $I_{AA}$  for Cu+Cu collisions 
at 200 GeV versus the number of
participant nucleons, $N_{part}$. Width of the away side jet is $\sigma$=0.22 radians. 
Circles are estimation for in plane, triangles - 
for out of plane. }
\end{figure}

Results of calculation 
for $I_{AA}$ in- and out of plane for Au+Au collisions at 200 GeV are presented in Fig.~\ref{fig:Iaa}. 
One can see the main feature of the surface biased scenario: in the most central events there are 
more di-jets out of the reaction plane than di-jets in the plane. This is opposite to the $R_{AA}$ reaction 
plane dependence or {\it to the scenario which assumes only``punch-through'' di-jets}. 
In mid-central events the situation is mixed, and in peripheral collisions 
there are more in-plane di-jets. Wide centrality bin of 20 to 60$\%$, commonly used in experiments, corresponds 
to $N_{part}$=125 if number of di-jets is weighted by the number of trigger particles. 
Resuts for Cu+Cu collisions are shown in  Fig.~\ref{fig:IaaCu}.

I want to emphasize that the role of event by event fluctuations should be treated by a better way 
than I do it here by parameter $a$. It is known that nucleon position fluctuations in the colliding Cu+Cu nuclei 
can significantly increase the observed $v_2$ in the most central collisions~\cite{manly}. This is why 
static models, including our model presented here, underestimate $v_2$ in the central Cu+Cu collisions. The same should be true 
when we consider di-jets in Au+Au or Cu+Cu: local density fluctuations induce positive correlation between 
surviving probability of the trigger jet and the away side jet. Again, such fluctuations could increase 
the observed value of $I_{AA}$, especially in the central collisions.

Few more comments: as I pointed out in the previous paper~\cite{paper2}, if we see a significant 
suppression of non-photonic electrons~\cite{electrons}, one can expect values of $v_2$ 
for such electrons at high $p_T$ to be equal to $v_2$ for high $p_T$ hadrons. Another point: 
suppression of the away 
side jets in direct gamma-hadron correlations, $I_{AA}$, should be equal to the $R_{AA}$ of 
inclusive hadrons at about gamma energy. Here I ignore the difference in quark and gluon 
fragmentation functions. The reason, that this may change the result, is that hadrons in the away side 
jets for
 the direct gamma-hadron correlations are mostly from quark fragmentation, while inclusive 
hadrons below 10 Gev/c are dominantly from gluon fragmentation.

All numbers for Au+Au collisions  are presented at 200 GeV in Table~\ref{table1}, at 62 GeV in Table~\ref{table2}. 
Numbers for Cu+Cu collisions at 200 GeV are in Table~\ref{table3}, and at 62 GeV are in Table~\ref{table4}.

\begin{table}[htb]
\caption{\label{table1}Results of calculations for Au+Au at 200 GeV. In all calculations 
a=0.2 fm/c and $\sigma$=0.22 radians.
}
\begin{center}
\begin{tabular}{cccccccccc}
cent, $\%$ & $N_{part}$ & R& $R_{in}$ &  $R_{out} $& $v_2$, $\%$ & $<I_{AA}>$ & $I_{in}$ & $I_{out}$ \\
\hline
0-5  & 353 & 0.195& 0.213 & 0.177 & 4.71 & 0.145 & 0.0882 & 0.202 \\
\hline
5-10 & 300 & 0.235 & 0.281 & 0.211 & 7.06 & 0.198 & 0.135 & 0.262 \\
\hline
10-15 & 255 & 0.29 & 0.348 & 0.246 & 8.64 & 0.259 & 0.204& 0.314\\
\hline
15-20 & 215 & 0.335 & 0.417 & 0.282 & 9.68 & 0.34& 0.317& 0.364\\
\hline
20-25 & 182 & 0.39 & 0.483 & 0.316 & 10.47 & 0.448& 0.488& 0.407 \\
\hline
25-30 & 152 & 0.45 & 0.563 & 0.362 & 10.87 & 0.575& 0.687& 0.463\\
\hline
30-35 & 126 & 0.505 & 0.632 & 0.40 & 11.23 & 0.659& 0.809& 0.508\\
\hline
35-40 & 103 & 0.565 & 0.705 & 0.447 & 11.19 & 0.727& 0.897& 0.557\\
\hline
40-45 & 84 & 0.635 & 0.779 & 0.512 & 10.37 & 0.783& 0.947& 0.62\\
\hline
45-50 & 67 & 0.715 & 0.848 & 0.589 & 9.03 & 0.833& 0.974& 0.69\\
\hline
50-55 & 52 & 0.79 & 0.902 & 0.681 & 7.0 &0.877&0.99&0.763\\
\hline
55-60 & 40 & 0.84 & 0.958 & 0.847 & 5.53 &&&\\
\hline
cent, $\%$ & $N_{part}$ & R& $R_{in}$ &  $R_{out}$ & $v_2$, $\%$ & $<I_{AA}>$ & $I_{in}$ & $I_{out}$ \\
\end{tabular}
\end{center}
\end{table}
 
\begin{table}[t]
\caption{\label{table2}Results of calculations for Au+Au at 62 GeV.}
\begin{center}
\begin{tabular}{cccccc}
cent, $\%$ & $N_{part}$ & R& $R_{in}$ &  $R_{out} $& $v_2$, $\%$ \\
\hline
0-5 & 345 & 0.333 & 0.364 & 0.302 & 4.65 \\
\hline
5-10 & 296 & 0.402 & 0.461 & 0.344 & 7.29 \\
\hline
10-15 & 250 & 0.468 & 0.553 & 0.383 & 9.06 \\
\hline
15-20 & 211 & 0.532 & 0.639  & 0.425 & 10. \\
\hline
20-25 & 178 & 0.591 & 0.719  & 0.462 & 10.9 \\
\hline
25-30 & 149 & 0.656 & 0.801 & 0.512 & 11.\\
\hline
30-35 & 123.8 & 0.707 & 0.861 & 0.553 & 10.9 \\
\hline
35-40 & 102 & 0.756 & 0.911 & 0.601 & 10.3 \\
\hline
40-45 & 82.9 &  0.805 & 0.948 & 0.663 & 8.83 \\
\hline
45-50 & 66.2 &  0.851 & 0.971 & 0.731 & 7.05 \\
\hline
50-55 & 51.9 &  0.895 & 0.984 & 0.805 & 5.01\\
\hline
55-60 & 40 &  0.922 & 0.99 & 0.854 & 3.69 \\
\hline
cent, $\%$ & $N_{part}$ & R& $R_{in}$ &  $R_{out} $& $v_2$, $\%$ \\
\end{tabular}
\end{center}
\end{table}

\begin{table}[htb]
\caption{\label{table3}Results of calculations for Cu+Cu at 200 GeV.}
\begin{center}
\begin{tabular}{ccccccccc}
cent, $\%$ & $N_{part}$ & R& $R_{in}$ &  $R_{out} $& $v_2$, $\%$ & $<I_{AA}>$ & $I_{in}$ & $I_{out}$ \\
\hline
0-5 & 110 & 0.37 &  0.40 & 0.34 & 3.4 & 0.194 & 0.136 & 0.251 \\
\hline
5-10 & 95 & 0.435 & 0.48 & 0.39 & 5.1 & 0.263 & 0.203 & 0.324 \\
\hline
10-15 & 82 &  0.51 & 0.57 & 0.45 & 6.0 & 0.346 & 0.31 & 0.382 \\
\hline
15-20 & 70.5 & 0.585 & 0.66 & 0.51 & 6.3 &  0.482&  0.497& 0.467 \\
\hline
20-25 & 60 & 0.655 & 0.74 & 0.57 & 6.4 & 0.581 & 0.637 & 0.526  \\
\hline
25-30 & 51 & 0.7 & 0.79 & 0.61 & 6.4 &  0.645& 0.723 & 0.568 \\
\hline
30-35 & 43 & 0.775 & 0.86 & 0.69 & 5.4 & 0.733 & 0.817 & 0.648\\
\hline
35-40 & 36 & 0.825 & 0.9 & 0.75 & 4.3 & 0.784 & 0.865 & 0.704\\
\hline
40-45 & 29 & 0.87 & 0.93 & 0.81 & 3.3 & 0.833 & 0.903 & 0.762 \\
\hline
45-50 & 24 & 0.91 & 0.95 & 0.87 & 2.1 &  &  &   \\
\hline
50-55 & 19 & 0.95 & 0.97 & 0.93 & 0.9 &  &  &   \\
\hline
55-60 & 15 & 0.98 & 0.98 & 0.98 & 0 &  &  &   \\
\hline
cent, $\%$ & $N_{part}$ & R& $R_{in}$ &  $R_{out} $& $v_2$, $\%$ & $<I_{AA}>$ & $I_{in}$ & $I_{out}$  \\
\end{tabular}
\end{center}
\end{table}

\begin{table}[htb]
\caption{\label{table4}Results of calculations for Cu+Cu at 62 GeV.
}
\begin{center}
\begin{tabular}{cccccc}
cent, $\%$ & $N_{part}$ & R& $R_{in}$ &  $R_{out} $& $v_2$, $\%$  \\
\hline
0-5 & 104 & 0.598 & 0.638 & 0.559 & 3.28   \\
\hline
5-10 & 92 & 0.672 & 0.737 & 0.607 & 4.82  \\
\hline
10-15 & 80 & 0.736 & 0.817 & 0.656 & 5.48  \\
\hline
15-20 & 68 & 0.8 & 0.888 & 0.712 & 5.48 \\
\hline
20-25 & 57 & 0.845 & 0.931 & 0.759 & 5.08 \\
\hline
25-30 & 48.6 & 0.874 & 0.954 & 0.793 & 4.61  \\
\hline
30-35 & 40.6 & 0.91 & 0.975 & 0.846 & 3.52 \\
\hline
35-40 & 34 & 0.932 & 0.982 & 0.882 & 2.7\\
\hline
40-45 & 28 & 0.952 & 0.989 & 0.915 & 1.93  \\
\hline
cent, $\%$ & $N_{part}$ & R& $R_{in}$ &  $R_{out} $& $v_2$, $\%$ \\
\end{tabular}
\end{center}
\end{table}

\end{document}